\documentstyle[12pt,twoside,fleqn,espcrc1]{article}
\input{psfig}

\title{\vspace*{-1cm}
\hfill MKPH-T-97-32\\
{\bf   Explicit pionic degrees of freedom in deuteron 
  photodisintegration in the $\Delta$-resonance region}
 \thanks{Supported by the Deutsche Forschungsgemeinschaft (SFB 201).}
 \thanks{Contribution to Few-Body XV, Groningen (1997)}
}

\author
{M.\ Schwamb$^{\,\,\mbox{\scriptsize{a}}}$, 
H.\ Arenh\"ovel$^{\,\,\mbox{\scriptsize{a}}}$, 
 P.\ Wilhelm \address{Institut f\"ur Kernphysik,           
  Johannes Gutenberg-Universit\"at, D-55099 Mainz, Germany}
and
Th.\ Wilbois \address{Institut f\"ur Theoretische Physik, Universit\"at Hannover,
 D-30167 Hannover, Germany}
}

\begin{document}
\maketitle

\begin{abstract}
Photodisintegration of the deuteron above $\pi$-threshold 
 is studied in a coupled
channel   approach including $N \Delta$- and  $\pi d$-channels  with
 pion retardation in potentials and exchange currents. 
\end{abstract}

\section{Introduction}
Photodisintegration of the deuteron in the $\Delta$-resonance region 
is particularly interesting in order to investigate 
 the $N \Delta$-interaction.
 None of the models developed so far 
 is able to describe in a satisfactory manner the experimental data  
 over  the whole $\Delta$-resonance region 
(for a review see \cite{ArS91}).
 Among the most sophisticated approaches  are  
 the  unitary three-body model of Tanabe and Ohta \cite{TaO89} 
and the coupled channel  approach (CC) of Wilhelm and Arenh\"ovel
\cite{WiA93}.  In both models, all  free parameters
 were fixed in advance  by fitting
 $NN$- and $\pi N$-scattering, and $\pi$-photoproduction on the nucleon. 
Consequently, no adjustable parameters remained for deuteron 
 photodisintegration. 
However, it turned out that both approaches 
 considerably underestimated the total cross 
section in the $\Delta$-region  by about 20-30\% \cite{{TaO89},{WiA93}}. 
Another failure was the wrong shape of the differential 
 cross section and the  photon asymmetry, 
especially at photon  energies above  300 MeV
   \cite{{TaO89},{WiA93},{Leg95}}.
 
\begin{figure}[htb]
\centerline{\psfig{figure=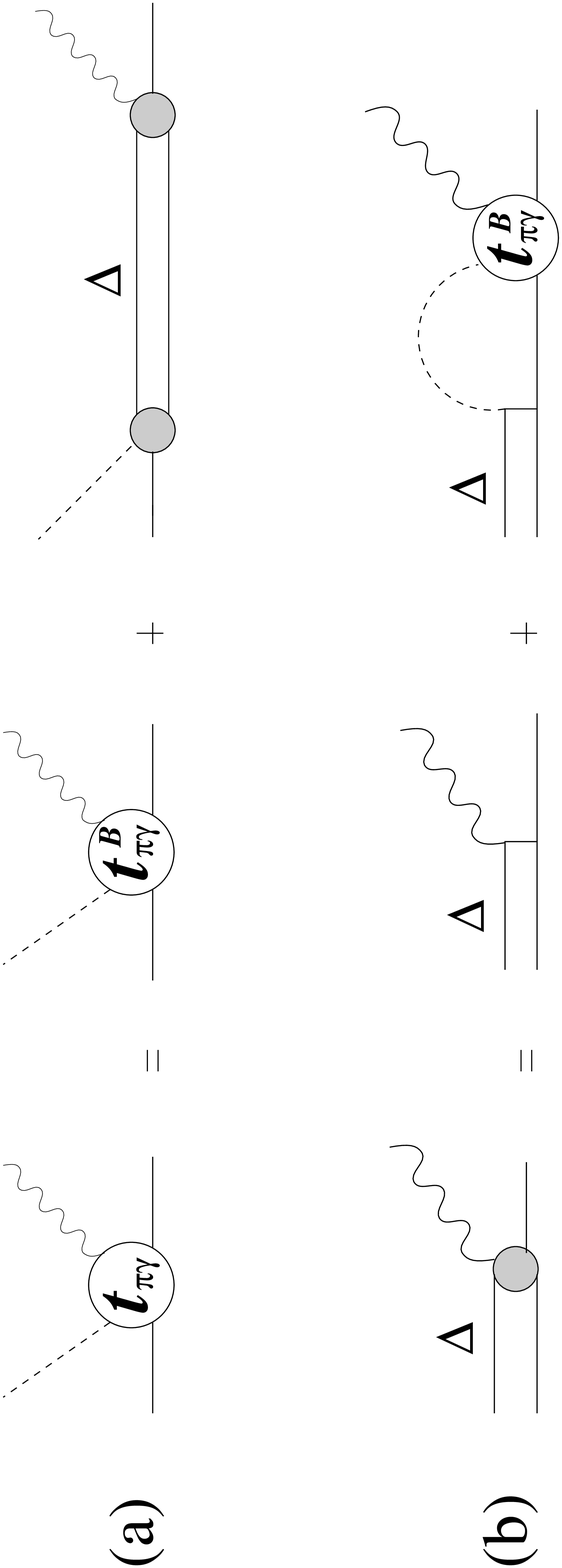,width=10cm,angle=270}}
\vspace{-2.2cm}
\caption{(a) The 
$M_{1+}(\frac{3}{2})$-multipole amplitude of pion photoproduction 
consisting of a Born  and a resonant amplitude.
 (b) The dressed $\gamma N \Delta$-coupling, including nonresonant 
 pion rescattering.}
\label{tmatrix}
\end{figure}

In these calculations,  one of the principal problems 
is the question of how to fix the 
$\gamma N \Delta$-coupling $G^{M1}_{\Delta N}(E_{\Delta})$ in the
 $M1 \,\,  N \Delta$-current 

\begin{equation}\label{ndcurrent}
\vec{\jmath}^{\,\,\, M1}_{\Delta N}(E_{\Delta}, \vec{k}) =
\frac{G^{M1}_{\Delta N}(E_{\Delta})}{2M} \,\, \tau_{\Delta N,0} \,\,
i\, \vec{ \sigma}_{\Delta N}
\times \vec{k} \, \, ,
\end{equation}
 where  $E_{\Delta}$ is the energy
 available for the internal excitation of the $\Delta$
 and $\vec{k}$  the momentum of the incoming photon.
 Wilhelm et al.\ as well as Tanabe et al.\  have determined
 $G^{M1}_{\Delta N}(E_{\Delta})$
  by fitting the $M_{1+}(3/2)$-multipole of pion 
 photoproduction on the nucleon.
 The full pion production amplitude $t_{\pi \gamma}(E_{\Delta})$ 
 in the $(3,3)$-channel can be written as
\begin{equation}\label{deltat}
t_{\pi \gamma}(E_{\Delta}) = t^{B}_{\pi \gamma}(E_{\Delta}) -
 \frac{v^{\dag}_{\Delta} \vec{\epsilon} 
\cdot \vec{\jmath}^{\,\,\, M1}_{\Delta N}(E_{\Delta},\vec{k})}{
      E_{\Delta}-M^{0}_{\Delta} -\Sigma_{\Delta}(E_{\Delta})} \,\, ,
\end{equation}
where $t^{B}_{\pi \gamma}(E_{\Delta})$ is the nonresonant Born amplitude.
While in \cite{WiA93} 
an effective 
$\gamma  N \Delta$-coupling  $G^{M1}_{\Delta N}(E_{\Delta}) $
 and the model of   \cite{PoS87} for 
 the bare $\Delta$-mass  $M^{0}_{\Delta}$, the $\Delta$-self energy 
 $\Sigma_{\Delta}(E_{\Delta})$ 
 and the $\Delta \pi N$-vertex  $v^{\dag}_{\Delta}$ 
has been used,
we follow here the work of Tanabe and Ohta (model A in \cite{TaO85}). 
$G^{M1}_{\Delta N}(E_{\Delta})$
contains besides  the bare $\gamma N \Delta$-coupling
 the contributions from  nonresonant pion rescattering
 (Fig.\ \ref{tmatrix}), so that it becomes complex and energy dependent.

\begin{figure}[htb]
\centerline{\psfig{figure=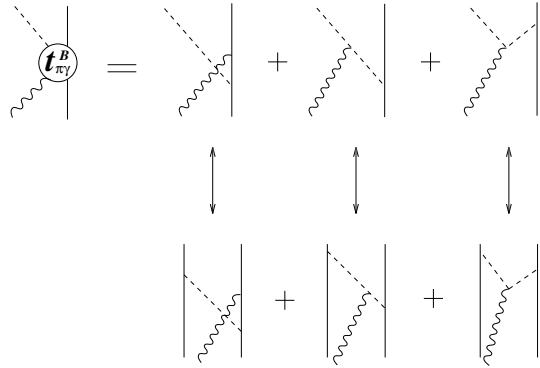,width=7cm,angle=270}}
\vspace{-.7cm}
\caption{The Born terms contributing to the 
$M_{1+}(\frac{3}{2})$-multipole amplitude of pion photoproduction 
(upper part) and their correspondence in the two-body 
 recoil and  meson currents (lower part).}
\label{vergleich}
\end{figure}

 The Born terms
 contributing to the $(3,3)$-channel are the crossed $N$-pole and 
 $\pi$-pole graphs. 
 When embedded into the two-nucleon system, these Born terms become part
of the two-body  recoil and the $\pi$-meson currents, respectively 
 (Fig.\ \ref{vergleich}).

In static calculations, however, the recoil current
  is not present due to its
 cancellation against the wave function renormalization current \cite{GaH76}. 
 A similar, but less  serious problem  arises in the treatment of
 the pion pole diagrams compared to the meson current of static MEC.
It had already been conjectured  in \cite{WiA93} that this inconsistent
 treatment
 of  pion exchange may lead to the observed underestimation of the total cross 
 section in their coupled channel approach, because by  incorporating 
 the Born terms effectively 
 into an increased   $M1\,\, \Delta$-excitation 
 strength, a satisfactory agreement with the data could be achieved.
 In order to avoid these shortcomings, we have  included for the first time 
 in a coupled channel approach  complete retardation in the $\pi$-exchange  
 contributions to potentials and  MECs.

\section{The Model}

 Concerning the potential models  which enter our coupled channel 
 approach, we have chosen for the retarded NN-potential an improved
 version of   the energy dependent Bonn-OBEPT   developed by Elster et al.\,
  \cite{MaH88}, which has to be renormalized via subtraction of a
 $N \Delta$-box graph \cite{PoS87}.
 Transitions between  $NN$- and $N \Delta$-space are mediated by 
 retarded $\pi$- and $\rho$-exchange whose form factors are 
 fixed  by fitting the $^1D_2$ 
 $NN$-partial wave. 
  In order to ensure unitarity up to the $2 \pi$-threshold, we  
 consider in addition the formation of an intermediate $NN$-state with 
 the quantum numbers of the deuteron 
 and   a pion as spectator (denoted for simplicity by $\pi d$-channel).

\begin{figure}[htb]
\centerline{\psfig{figure=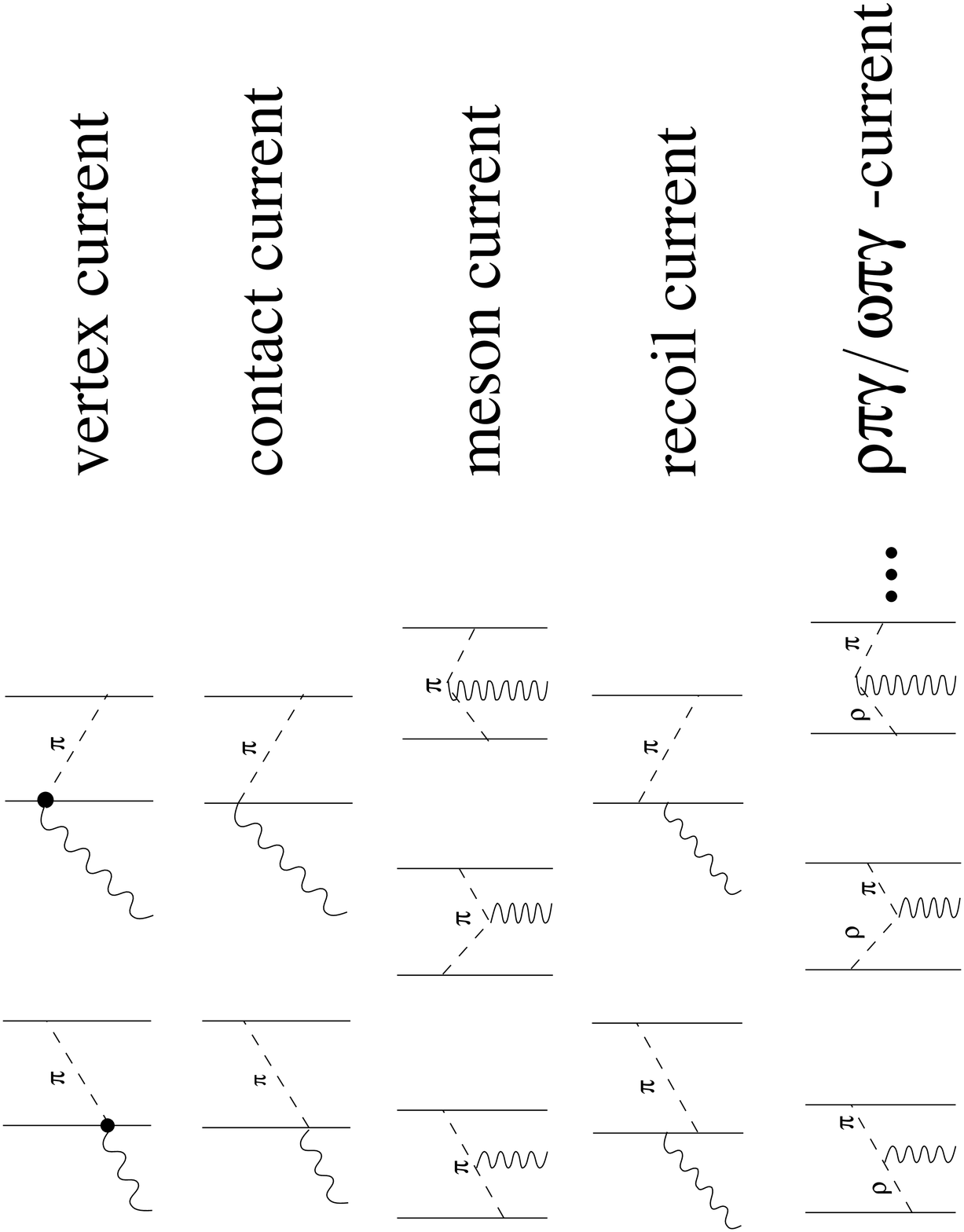,width=10cm,angle=270}}
\vspace{-.7cm}
\caption{Graphical representation of the retarded $\pi$-MECs.}
\label{mecdarstellung}
\end{figure}

  Concerning the e.m.\ part of our model, 
 the  $\Delta$-excitation is the most important photoabsorption mechanism
  above $\pi$-threshold.
 It  is described by the current operator in Eq.\ (\ref{ndcurrent})
 neglecting small E2 contributions.
Concerning gauge invariance, we are able to  show  that
 current conservation for the $\pi$-retarded MECs
 is fulfilled if we consider besides the usual vertex-, meson- and 
contact-MECs the recoil current, the recoil 
  and additional two-body  charge densities
 (Fig.\ \ref{mecdarstellung}).

Whereas the effect of the additional two-body charge terms is very small,
 the recoil contributions  turn out to 
 be quite important (see discussion below).
 They do  not appear in static approaches 
 due to their  cancellation against the  wave function renormalization 
 contributions \cite{GaH76}, which have  
their  origin  in the renormalization of the baryonic states
 when eliminating the mesonic wave function components.
This concept breaks down  beyond the $\pi$-threshold 
 if full $\pi$-retardation is considered since the  $\pi N N $-component 
can be on-shell.  Therefore, we do not orthonormalize and no
wave function renormalization contributions  appear.
Consequently, the recoil current and   charge densities  
have to be included. Because 
 the pion production model of Tanabe and Ohta {\cite{TaO85} effectively
incorporates $\omega$-exchange, we include in addition the leading order
 $\rho \pi \gamma$- and $\omega \pi \gamma$-currents, which are purely 
transverse \cite{RiG97}.
Because  the  $\rho$-mass  is rather large, 
retardation in the $\rho$-MEC is expected to be rather unimportant 
 and therefore not considered in this work.
 
\section{Results}
Our results for the total photodisintegration cross section are shown in
Fig.\ \ref{sigtot}. Similar to  \cite{WiA93},
the static calculation considerably underestimates the data.
Inclusion of retardation in the hadronic interaction even lowers further the 
cross section, which is more than compensated by  retardation in 
 the $\pi$-MEC leading to a strong  enhancement
 which can be traced back essentially to the inclusion of 
recoil contributions. The inclusion of the $\pi d$-channel 
 and the $\rho \pi \gamma / \omega \pi \gamma$-MECs  enhances the
 cross section further so that our full calculation now  gives 
 quite a good agreement with  experimental data over the whole energy range.
 In Figs.\ \ref{wqdiff} and \ref{sigma}, we show differential cross sections
 and photon asymmetries for various energies. Whereas the differential cross 
 section is in satisfactory agreement with the data, we slightly underestimate
 the absolute size of the asymmetry. However, in contrast to 
 \cite{{TaO89},{WiA93}} we are able to reproduce quite well 
 the shape of these two  observables at higher energies.

\vspace{0.5cm}
\centerline{\bf ACKNOWLEDGMENT}
We would like to thank S.\ Wartenberg from the A2 collaboration for providing
 us the  preliminary results on the photon asymmetry prior to
 publication \cite{War97}.

\begin{figure}[htp]
\centerline{\psfig{figure=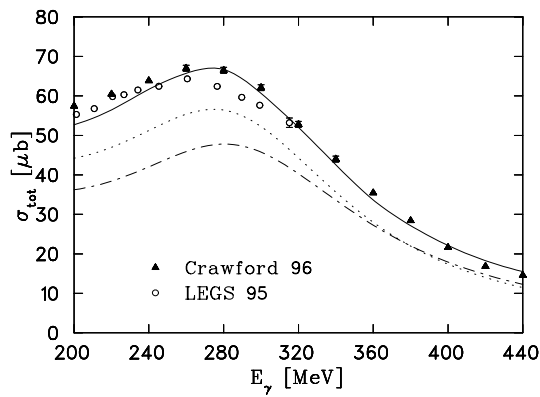,width=7 cm,angle=90}}
\vspace{-.7cm}
\caption{Total cross section for $\gamma d \rightarrow p n$ as a function
of photon laboratory energy $E_{\gamma}$ in comparison with experiment 
{\protect \cite{{Leg95},{Cra96}}}.
Dotted: static OBEPR-calculation, 
dash-dot: retardation switched on only in the
 hadronic part but  static MECs ,
full: calculation with complete retardation, 
 $\pi d$-channel and  $\rho \pi \gamma / \omega \pi \gamma$-MECs.}
\label{sigtot}
\end{figure}

\begin{figure}[htb]
\centerline{\psfig{figure=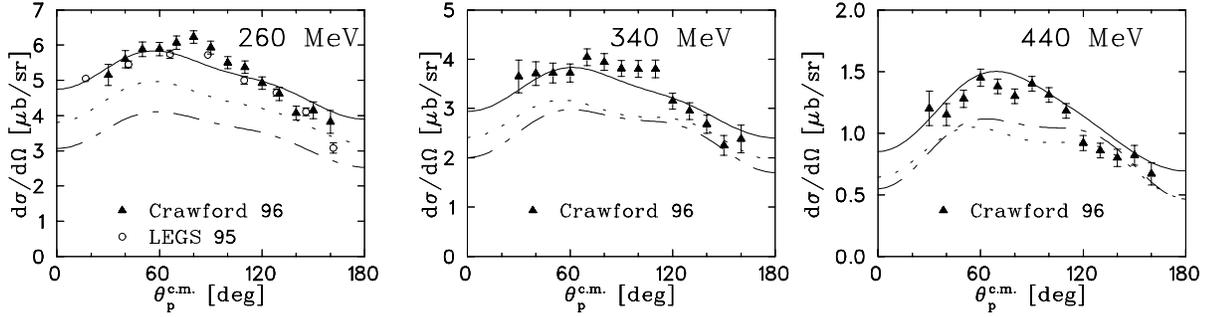,width=15.8cm,angle=90}}
\vspace{-.7cm}
\caption{Differential cross section for various energies in comparison with 
         experiment {\protect \cite{{Leg95},{Cra96}}}.
          Notation of the curves as in Fig.\ {\protect \ref{sigtot}}.}  
\label{wqdiff}
\end{figure}

\begin{figure}[htb]
\centerline{\psfig{figure=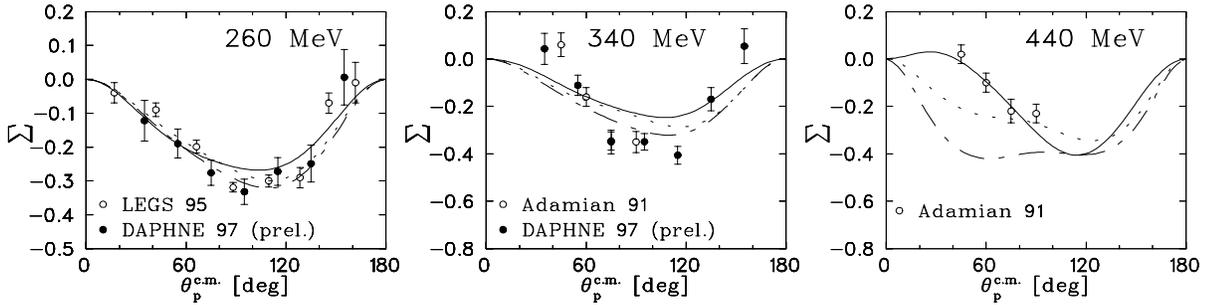,width=15.8cm,angle=90}}
\vspace{-.7cm}
\caption{Photon asymmetry $\Sigma$ for various energies in comparison with 
         experiment {\protect \cite{{Leg95},{War97},{Ada91}}}.
          Notation of the curves as in 
         Fig.\ {\protect \ref{sigtot}}.}  
\label{sigma}
\end{figure}


\begin{thebibliography}{888.}
\bibitem{ArS91}
H.\ Arenh\"ovel and M.\ Sanzone,  Few-Body Syst.\ Suppl.\ 3 (1991) 1
\bibitem{TaO89}
H.\ Tanabe and K.\ Ohta, Phys.\ Rev.\ C 40 (1989) 1905
\bibitem{WiA93}
P.\ Wilhelm and H.\  Arenh\"ovel, Phys.\ Lett.\  B 318 (1993) 410
\bibitem{Leg95}
The LEGS Collaboration, Phys.\ Rev. C 52 (1995) R455
\bibitem{PoS87}
H.\ P\"opping,  P.\ U.\ Sauer and X.-Z.\ Zang, Nucl.\ Phys.  A 474 (1987) 557\\
H.\ P\"opping,  P.\ U.\ Sauer and X.-Z.\ Zang, Nucl.\ Phys.  A 550 (1992) 563
\bibitem{TaO85}
H.\ Tanabe and K.\ Ohta, Phys.\ Rev.\  C 31 (1985) 1876
\bibitem{GaH76}
M.\ Gari and H.\  Hyuga, Z.\ Phys.\  A 277 (1976) 291
\bibitem{MaH88}
Ch.\ Elster,  W.\ Ferchl\"ander, K.\ Holinde, D.\ Sch\"utte and R.\ Machleidt, 
 Phys.\ Rev.\  C 37 (1988) 1647
\bibitem{RiG97}
F.\ Ritz, H.\  G\"oller, Th.\ Wilbois and H.\  Arenh\"ovel,
Phys.\ Rev.\  C 55 (1997) 2214
\bibitem{Cra96}
  R.\ Crawford et al., Nucl.\ Phys.\ A 603 (1996)  303
\bibitem{War97}
S.\ Wartenberg (A2 collaboration), private communication, Mainz 1997
\bibitem{Ada91}
F.\ V.\ Adamian et al., J.\ Phys.\ G.\  17 (1991) 1189 
\end{thebibliography}
\end{document}